# The $\beta$ function and equation of state for QCD with two flavors of quarks

T. Blum, Leo Kärkkäinen and D. Toussaint

*Department of Physics, University of Arizona,*

*Tucson, AZ 85721, USA*

Steven Gottlieb

*Department of Physics, Indiana University,*

*Bloomington, IN 47405, USA*

(October 21, 1994)

## Abstract

We measure the pressure and energy density of two flavor QCD in a wide range of quark masses and temperatures. The pressure is obtained from an integral over the average plaquette or $\langle\bar\psi\psi\rangle$. We measure the QCD $\beta$ function, including the anomalous dimension of the quark mass, in new Monte Carlo simulations and from results in the literature. We use it to find the interaction measure, $\varepsilon - 3p$, yielding non-perturbative values for both the energy density $\varepsilon$ and the pressure $p$.
12.38.Gc, 11.15.Ha

# I. INTRODUCTION

We expect that at high temperatures strong interactions will enter a new phase, the quark-gluon plasma (QGP), which is believed to have existed in the extremely high temperatures microseconds after the big bang. Heavy-ion collision experiments at the Brookhaven AGS and CERN are currently trying to recreate the QGP. In order to prove its existence in the aftermath of a heavy-ion collision and to understand the dynamics of the QGP in the early universe one needs as input, among other things, the equation of state for the system. Since the phase transition occurs in a regime of strong gauge coupling, a non-perturbative method is called for. Lattice calculations provide such a method. However, at currently practical lattice spacings the operator formalism usually used in such calculations for the equation of state is not necessarily non-perturbative, since it requires the knowledge of the asymmetry coefficients, or Karsch coefficients [1]. These are currently known only perturbatively [2]. These asymmetry coefficients are short distance quantities defined at the scale of the lattice spacing $a$, so that if $a$ could be made small enough (and temporal size $N_t$ large enough to keep the temperature fixed) the perturbative coefficients could be used accurately even though the temperature would remain at a typical QCD length scale. However, the use of perturbative values for the asymmetry coefficients leads to distortions in the equation of state at the bare couplings used today. An effort can be made to non-perturbatively measure the asymmetry coefficients [3]. In practice that has turned out to be difficult [4].

The integral method does not require the knowledge of these coefficients. It was first used in the context of lattice QCD to calculate the interface tension by S. Huang *et al.* [5] and later modified for the bulk pressure of pure gauge QCD by J. Engels *et al.* [6]. The disadvantage of the integral method is that for the pressure at a single temperature and quark mass, a number of different simulations are required in order to provide the integrand.

We have done an extensive survey of the gauge coupling and quark mass plane for two flavor QCD, allowing us to measure the non-perturbative pressure by integration. Using data from the literature for the $\rho$ and $\pi$ meson masses, we calculate the non-perturbative $\beta$ function to compute the interaction measure and hence the energy density.

Although these simulations avoid one of the problems of using QCD simulations on small lattices, namely the use of perturbation theory, the problems of scaling violation, or non-constant ratios of physical lengths, of flavor symmetry breaking with the Kogut-Susskind quarks, and of effects on the thermodynamics of replacing integrals over the momenta by sums over discrete Matsubara frequencies remain.

In Sec. II we present the formalism for calculation of thermodynamic quantities and the $\beta$-function. Section III details our simulations and the results for the energy and pressure.

## II. THEORY

A Euclidean $N_s^3 \times N_t$ lattice with periodic boundary conditions has a temperature $T$ and volume $V$ given by

$$V = N_s^3 a^3,$$
$$1/T = N_t a \ , \tag{1}$$

where $a$ is the lattice spacing. The form of the partition function $Z$ for QCD with $n_f$ fermions with Kogut-Susskind (KS) discretization is

$$Z = \int [dU_{(n,\mu)}] \exp\{ (6/g^2)S_g + (n_f/4) \operatorname{Tr} \log[am_q + \not{D}] \} \ , \tag{2}$$

where the gauge fields $U_{(n,\mu)}$ are SU(3) matrices at the link $(n,\mu)$ (at site $n$, to the direction $\mu = 0,...,3$), $(6/g^2)$ is the gauge coupling and $am_q$ the bare quark mass in lattice units. The gauge action

$$S_g = \frac{1}{3}\operatorname{Re} \sum_{n,\mu<\nu} \operatorname{Tr} U_\square(n,\mu,\nu) \ , \tag{3}$$

is a function of $U_\square(n,\mu,\nu)$, the path ordered product of link matrices around the elementary plaquette at site $n$ in the $\mu\nu$ plane. The covariant derivative $\not{D}$ contains the Kogut-Susskind phases. For $n_f = 2$, the simulation is performed using the standard refreshed molecular dynamics algorithm [7]. This involves integration of an equation of motion with a nonzero step, and a resulting error in physical averages. As it turns out, this step size error must be handled with care.

### A. Thermodynamics

Thermodynamic variables are derivatives of the partition function $Z$ defined in Eq. (2). In particular, the pressure $p$ and energy density $\varepsilon$ are given by

$$\varepsilon V = -\frac{\partial \log Z}{\partial (1/T)} \tag{4}$$

and

$$\frac{p}{T} = \frac{\partial \log Z}{\partial V} \quad . \tag{5}$$

For large, homogeneous systems the free energy is proportional to the volume:

$$\log Z = V \frac{\partial \log Z}{\partial V} \quad . \tag{6}$$

Thus, the free energy density $f$ can be connected to the pressure by

$$\frac{pV}{T} = \log Z = \frac{-fV}{T} \quad . \tag{7}$$

The free energy cannot be obtained from a single simulation, but its derivatives can be. In particular,

$$\langle \Box \rangle = \frac{-1}{2 N_s^3 N_t} \frac{\partial \log Z}{\partial (6/g^2)} \quad , \tag{8}$$

where $\langle \Box \rangle$ is the average plaquette normalized to 3 for a lattice of unit matrices and

$$\langle \bar{\psi}\psi \rangle = \frac{-1}{N_s^3 N_t} \frac{\partial \log Z}{\partial (am_q)} \quad . \tag{9}$$

These are relatively easy to measure in a simulation. If a series of runs is performed with different $am_q$ and $6/g^2$ values the pressure can be obtained by numerically integrating Eqs. (8) and (9). With the plaquette we obtain

$$\frac{pV}{T}(6/g^2, am_q) = 2N_s^3 N_t \int_{\text{cold}}^{6/g^2} [\langle \Box(6/g'^2, am_q) \rangle - \langle \Box(6/g'^2, am_q) \rangle_{\text{sym}}] d(6/g'^2) \ , \qquad (10)$$

where $\langle \Box(6/g^2) \rangle_{\text{sym}}$ is the average plaquette from a symmetric (cold) system. The subtraction of the symmetric value removes the divergent zero temperature pressure. The lower limit for the integration should be in a region where the difference between the zero and nonzero temperature plaquette expectation values is negligible. This removes the unknown constant introduced by the integration. In a completely analogous way we get from $\langle \bar{\psi}\psi \rangle$:

$$\frac{pV}{T}(6/g^2, am_q) = N_s^3 N_t \int_{\text{cold}}^{am_q} [\langle \bar{\psi}\psi(6/g^2, m_q'a) \rangle - \langle \bar{\psi}\psi(6/g^2, m_q'a) \rangle_{\text{sym}}] d(m_q'a) \ . \qquad (11)$$

or

$$\frac{p}{T^4}(am_q) = N_t^4 \int_{\text{cold}}^{am_q} [\langle \bar{\psi}\psi(m_q'a) \rangle - \langle \bar{\psi}\psi(m_q'a) \rangle_{\text{sym}}] d(m_q'a) \ . \qquad (12)$$

At high temperatures, $p/T^4$ should approach a constant. This means that the integrand in Eq. 10 must approach zero at large $6/g^2$. Figure 1 shows the $N_t = 4$ plaquettes and cold lattice plaquettes as a function of $6/g^2$ at $am_q = 0.1$. As required, the curves join at large $6/g^2$ values.

Although the average plaquette curves coalesce, the difference between the spatial and temporal plaquettes approaches a constant as shown in Fig. 2, and as required by the operator formula for the entropy. This means that at very high temperatures the spatial and temporal plaquettes are shifted by equal amounts but in opposite directions from the zero temperature plaquette.

### B. Energy density

To obtain the equation of state for QCD we also need the energy density $\varepsilon$. It can be obtained non-perturbatively using the $\beta$-function and the interaction measure $I$. The interaction measure $I$ is

$$\frac{IV}{T} = \frac{\varepsilon V}{T} - 3\frac{pV}{T} = -\frac{1}{T}\frac{\partial \log Z}{\partial (1/T)} + 3V\frac{\partial \log Z}{\partial V}$$

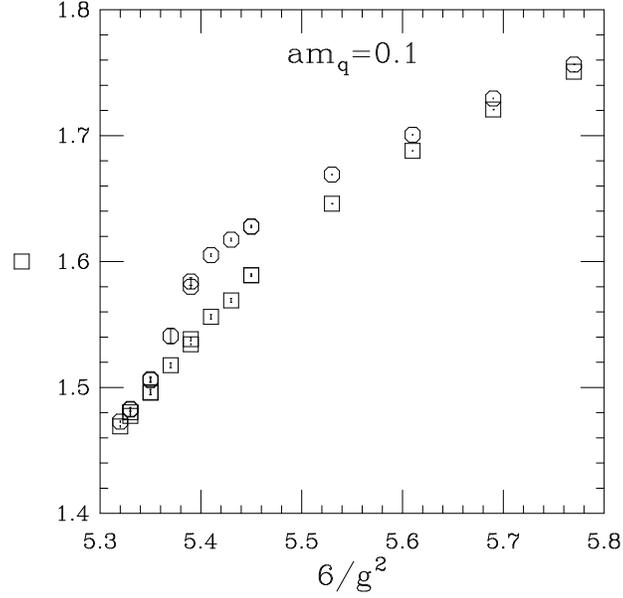

FIG. 1. The $N_t = 4$ plaquettes (octagons) and the symmetric plaquettes (squares) as a function of $6/g^2$ at $am_q = 0.1$.

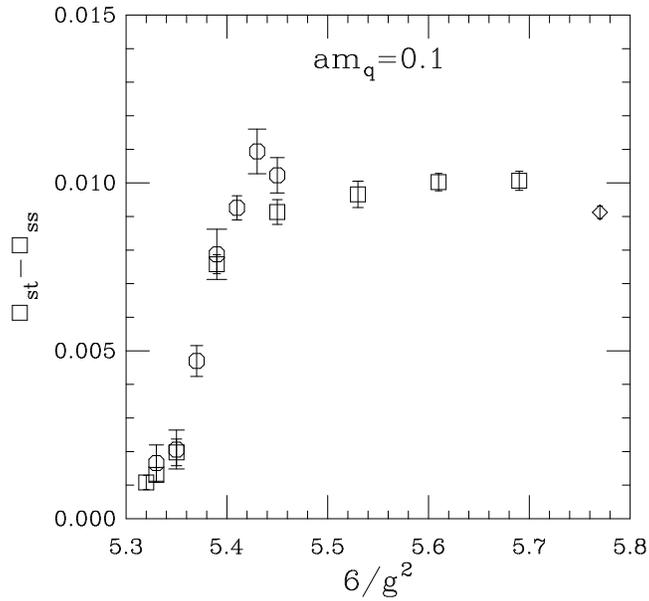

FIG. 2. The difference of spatial and temporal plaquettes as a function of $6/g^2$ at $am_q = 0.1$. The diamond corresponds to $16^3 \times 4$ lattice, the squares to $12^3 \times 4$ and the octagons to $8^3 \times 4$ lattices.

$$= (-a_t \frac{\partial}{\partial a_t} - 3a_s \frac{\partial}{\partial a_s}) \log Z = \frac{\partial \log Z}{\partial \log a}$$

$$= 2N_s^3 N_t \frac{\partial (6/g^2)}{\partial \log a}[\langle \Box \rangle - \langle \Box \rangle_{\text{sym}}] + N_s^3 N_t \frac{\partial (am_q)}{\partial \log a}[\langle \bar{\psi}\psi \rangle - \langle \bar{\psi}\psi \rangle_{\text{sym}}] \quad , \tag{13}$$

where $a_s$ and $a_t$ are the spatial and temporal lattice spacings. In the last step we have subtracted the zero temperature value. We have measured $\langle \Box \rangle$ and $\langle \bar{\psi}\psi \rangle$ on the lattice and calculated the $\beta$-function from mass spectrum data in the literature.

Knowledge of the non-perturbative pressure and interaction measure allows us to compute other bulk quantities. The energy and entropy $s$ become

$$\varepsilon = I + 3p \tag{14}$$

$$sT = I + 4p \quad . \tag{15}$$

Given the need for the $\beta$ function to find the interaction measure from Eq. 13, or for the asymmetry coefficients to find the energy directly, it would be nice to compute the energy density by an integration similar to that used for the pressure. To the extent that we can make a small change in temperature by changing $N_t$, this is possible. Let us start from Equation (4)

$$\varepsilon V = -\frac{\partial \log Z}{\partial (1/T)} \approx \frac{\Delta \log Z}{\Delta (N_t a)}$$

$$= \frac{N_s^3}{(N_t' - N_t)a} \int_{\text{cold}}^{am_q} N_t'[-\langle \bar{\psi}\psi \rangle_{N_t'} + \langle \bar{\psi}\psi \rangle_{\text{sym}}] + N_t[\langle \bar{\psi}\psi \rangle_{N_t} - \langle \bar{\psi}\psi \rangle_{\text{sym}}] d(m_q' a) \quad , \tag{16}$$

where $N_t'$ and $N_t$ are two different temporal extents. Of course, an analogous formula in terms of the plaquette also exists.

Equation (16) can be given in a form

$$\frac{\varepsilon}{T^4} \approx \frac{[(N_t' + N_t)/2]^4}{N_t' - N_t} \int_{\text{cold}}^{am_q} [N_t \langle \bar{\psi}\psi \rangle_{N_t} - N_t' \langle \bar{\psi}\psi \rangle_{N_t'} + (N_t' - N_t) \langle \bar{\psi}\psi \rangle_{\text{sym}}] d(m_q' a), \tag{17}$$

where we have taken as the inverse temperature the average $(N_t' + N_t)a/2$.

Ideally, we would like to take $N_t'a$ and $N_t a$ as close as possible to each other to increase the accuracy of the approximation of the derivative by the finite difference. This approximation gives a curve that is smoother than the real energy density. In fact, it gives the average $\varepsilon/T^4$ in the temperature range $(1/(N'_t a), 1/(N_t a))$. For linear regimes of $p/T^4$ it is exact. This is true, for example, in high temperatures, where the Boltzmann law is expected to be valid.

Approximating the pressure at $1/T = (N'_t + N_t)a/2$ with the average of $1/T' = N'_t a$ and $1/T'' = N_t a$ systems, we can get a formula for the entropy $s$ as well:

$$\frac{s}{T^3} \approx \frac{[N'_t N_t (N'_t + N_t)/2]^3}{N'^4_t - N^4_t} \int_{\text{cold}}^{am_q} [\langle \bar\psi\psi \rangle_{N_t} - \langle \bar\psi\psi \rangle_{N'_t}] dm'_q a. \tag{18}$$

The entropy does not need a vacuum subtraction; the symmetric average cancels out.

In the applications of the equation of state, the sound velocity of the thermal system is an important quantity. Acoustic perturbations travel in the system with a speed $c_s$:

$$\frac{1}{c_s^2} = \frac{d\varepsilon}{dp}. \tag{19}$$

One has to take the derivative keeping the physical quark mass fixed, *i.e.*, on the line of constant physics. Unfortunately, we know best only the variations of the energy density and pressure along lines of constant bare parameters. In order to measure the correct sound velocity one has to use the $\beta$ function to map the changes in bare parameters to physical changes of temperature and quark mass.

Define

$$\bar\varepsilon = \varepsilon a^4$$
$$\bar p = p a^4$$
$$\bar I = I a^4, \tag{20}$$

as the dimensionless quantities measured on the lattice. Then, the sound velocity becomes

$$\frac{1}{c_s^2} = 3 + \frac{d\bar I - 4\bar I \frac{da}{a}}{d\bar p - 4\bar p \frac{da}{a}}, \tag{21}$$

where the differentials have to be taken along a path which keeps $m_\pi/m_\rho$ constant. Therefore, it is not enough to have the energy density and pressure as the function of bare parameters; one needs the $\beta$ function as well. We will return to this in future work.

## C. The $\beta$-function

To change the lattice spacing keeping physical ratios fixed, we must adjust both relevant couplings in the action, $6/g^2$ and $am_q$, so our $\beta$-function has two components:

$$\beta(6/g^2, am_q) = \left( \frac{\partial (6/g^2)}{\partial \log a}, \frac{\partial (am_q)}{\partial \log a} \right), \tag{22}$$

where $a$ is the lattice spacing. This can be measured from $m_\pi$ and $m_\rho$ in units of the lattice spacing as functions of $6/g^2$ and $am_q$. In general, a change in the bare quantities changes both the lattice spacing $a$ and the physical quark mass. In practice, to keep physics constant while changing $a$, the partial derivatives in Eq. 22 are taken at constant $m_\pi/m_\rho$. Clearly, the nucleon mass or any other physical mass could be substituted for $m_\rho$, and in the continuum limit should give the same answer. In parctice, $m_\pi$ is special since it is uniquely sensitive to the quark mass. (In principle, other quantities sensitive to $am_q$ could be used, such as the nucleon-delta mass splitting.)

We have used $\pi$ and $\rho$ masses from simulations with two flavors of Kogut-Susskind fermions reported in the literature (see Table I) to calculate the $\beta$ function over a wide range of coupling and quark mass. A $2 \times 2$ matrix $M$ was formed by calculating derivatives from three or four points in coupling-quark mass space:

$$\begin{bmatrix} d\,\ln(am_\pi) \\ d\,\ln(am_\rho) \end{bmatrix} = \begin{bmatrix} \frac{\partial \ln(am_\pi)}{\partial (6/g^2)} & \frac{\partial \ln(am_\pi)}{\partial (am_q)} \\ \frac{\partial \ln(am_\rho)}{\partial (6/g^2)} & \frac{\partial \ln(am_\rho)}{\partial (am_q)} \end{bmatrix} \times \begin{bmatrix} d\,(6/g^2) \\ d\,(am_q) \end{bmatrix}. \tag{23}$$

The partial derivatives were found by linear interpolations among the points. The $\beta$ function was then calculated by inverting Eq. 23 with the changes on the left hand side set equal: $d\,\ln(am_\pi) = d\,\ln(am_\rho) = \delta$. This keeps $m_\pi/m_\rho$ constant. We have also calculated the $\beta$ function at $6/g^2 = 5.35$ and $am_q = 0.1$ from new simulations [9]. In these simulations, we varied space and time couplings anisotropically to measure the asymmetry coefficients. The asymmetry coefficients give the change in couplings as the space and time lattice spacings are adjusted independently. However, their sums give the symmetric change in the couplings when all lattice spacings are varied together, or the usual $\beta$ function. Values for the nonperturbative $\beta$ function are given in Table II, and a plot of the renormalization group (RG) flow is shown in Fig. 3. The values of $6/g^2$ and $am_q$ quoted in Table II are averages of the points used in calculating the $\beta$ function. In Table II we also give the value of the quark mass component of the $\beta$ function minus the contribution from the classical dependence on the lattice spacing. The RG flow depicted in Fig. 3 shows how to approach the continuum limit. From Table II we see that the gauge piece of the $\beta$ function is roughly half the perturbative value at couplings for the range of $6/g^2$ used in these simulations.

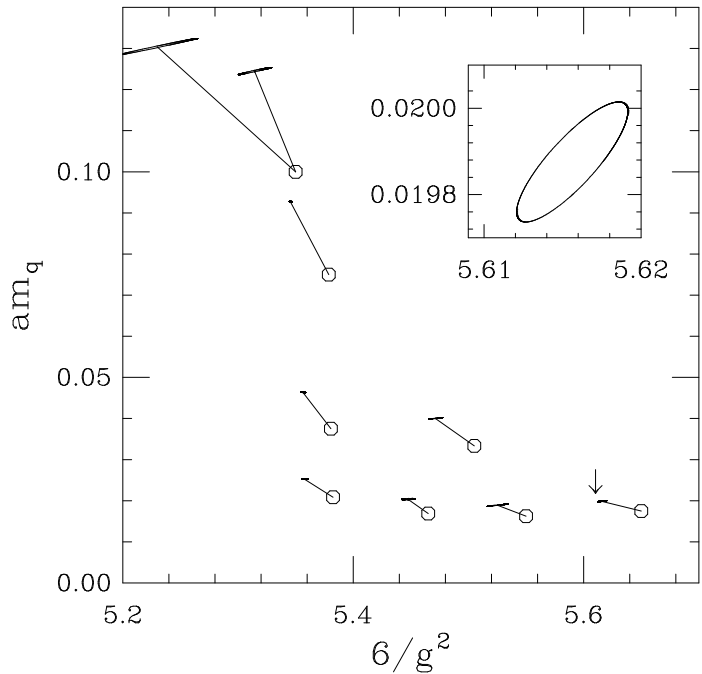

FIG. 3. Renormalization group flow in the gauge coupling-quark mass plane for two flavors of Kogut-Susskind fermions. The base of each line is indicated by an octagon and is where the $\beta$ function is evaluated. The end is indicated by a one standard deviation error ellipse. The inset shows this error ellipse for the $\beta$ function at $6/g^2 = 5.65$ and $am_q = 0.0175$, the point indicated by the arrow. The length of each line and its error ellipse corresponds to a scale change of $d\ln(a) = 0.1$, so this is actually 0.1 times the error on the $\beta$ function. The two lines at $6/g^2 = 5.35$ correspond to the $\beta$ function calculated using the VT $\rho$ and PV $\rho$ masses. (The larger line is for the VT $\rho$.) The differences are due to flavor symmetry breaking from the large lattice spacing.

The errors shown in Fig. 3 were calculated by forming a singular covariance matrix for the two components of the $\beta$ function for each of the hadron masses $am_h$ used in the calculation:

$$\begin{bmatrix} \left(\frac{\partial^2(6/g^2)}{\partial\ln(a)\partial(am_h)}\right)^2 & \frac{\partial^2(6/g^2)}{\partial\ln(a)\partial(am_h)}\frac{\partial^2(am_q)}{\partial\ln(a)\partial(am_h)} \\ \frac{\partial^2(6/g^2)}{\partial\ln(a)\partial(am_h)}\frac{\partial^2(am_q)}{\partial\ln(a)\partial(am_h)} & \left(\frac{\partial^2(am_q)}{\partial\ln(a)\partial(am_h)}\right)^2 \end{bmatrix} (\Delta am_h)^2 \qquad (24)$$

These matrices were added together to obtain a nonsingular covariance matrix. The covariance matrices for the first two entries in Table I were calculated from a jackknife estimate. The covariance matrices are diagonalized, and the allowed variance is then given by an ellipse whose semimajor and semiminor axes are along the eigenvectors of the covariance matrix and whose lengths are given by the square roots of the corresponding eigenvalues. The ellipses shown in Fig. 3 were calculated for one standard deviation. The error ellipses in Fig. 3 are close to straight lines, which indicates the two components of the $\beta$ function are highly correlated. Correlations between the $\pi$ and $\rho$ masses from the same simulations were not included in our analysis except for the point at $6/g^2 = 5.35$.

There are two systematic errors in our calculation of the $\beta$ function. The first comes from the linear approximation of the matrix of derivatives. However, Fig. 3 shows that the $\beta$ function is smooth over a wide range in coupling-quark mass space. This indicates the first order approximation is good even for these large changes. The second and most apparent systematic error comes from scaling violations in the masses (see 3). The point at $6/g^2 = 5.35$ and $am_q = 0.1$ was calculated using both the VT $\rho$ and the PV $\rho$ masses, and each gives a different answer for the $\beta$ function. (See Ref. [8] for a discussion of the meson operators.) The difference is attributable to the slopes of the masses as functions of $6/g^2$ and $am_q$. Since the VT $\rho$ and PV $\rho$ are not degenerate at this point but should be in the continuum limit, this is expected. At this coupling the slopes differ roughly by a factor of two, which leads to the large discrepancy in the $\beta$ function. At larger $6/g^2$ and smaller $am_q$ the problem is alleviated since the $\rho$ masses become degenerate to good accuracy.

Since we are interested in the $\beta$ function at many points in coupling - quark mass space to evaluate the interaction measure, a better method is to fit the spectrum data as a function of $6/g^2$ and $am_q$. Such a fit can also be used to transform functions of the bare quantities to functions of physical parameters like the temperature. Thus, we determine $m_\pi/m_\rho$ and $m_\rho a$ as functions of $am_q$ and $6/g^2$. The inverse function then yields the $\beta$ function. The fitting functions are given in Table III and shown in Fig. 4 where we have used $m_\rho = 0.770$ GeV to convert to inverse lattice spacing. These fitting forms are *ad hoc* fits to the masses in the relevant parameter region, and do not have the correct asymptotic behavior for large $6/g^2$ or large $am_q$. We take the functional form of $m_\pi/m_\rho$ from chiral perturbation theory with coefficients that are polynomials in $6/g^2$. We fit the mass ratio in this case because we could not obtain a fit with reasonable $\chi^2$ for the pion mass alone. These fitting functions are compared to the simulation results for $am_\pi$ and $am_\rho$ in Table I. For the region of $6/g^2$ relevant to the $N_t = 4$ thermal crossover, it can be seen that this is a good interpolating function for these masses. The $m_{\pi_2}$ results (not shown) are also fit reasonably well by a quadratic function of $6/g^2$ and $am_q$. In the continuum limit, the functions in Figs. 4(a) and 4(b) should converge, *i.e.*, $m_\pi = m_{\pi_2}$. However, at the couplings used in current lattice simulations, a pronounced breaking of the continuum $SU(2) \times SU(2)$ chiral symmetry is evident.

The $\beta$ function is determined by computing numerical derivatives of the bare parameters with respect to lattice spacing along lines of constant $m_\pi/m_\rho$, *i.e.*, from the inverse of the functions shown in Fig. 4. The confidence levels in Table III are low, which may indicate that the errors on the spectrum data are underestimated. In any case, the fits should be considered as giving smooth interpolated mass values in the parameter regions where data exist. Evidence that they work is the agreement between the resulting $\beta$ function and that from the direct calculation as shown in Fig. 5. We note that there seems to be a discrepancy at $6/g^2 = 5.35$ and $am_q = 0.1$. This is where the $\beta$ function was calculated from new simulations. These simulations show that the $\rho$ mass is a much flatter function of the bare parameters. As the quark mass is reduced below 0.1, it steepens appreciably. Since we have no spectrum data between $am_q = 0.1$ and $am_q = 0.05$, our fits can not resolve this behavior. In Table II we also give the $\beta$ function from the fitted spectrum at a few selected points, including extrapolations to zero quark mass.

The errors on the $\beta$ function from the fitted spectrum were calculated as described above for the direct method. That is, we determined a singular covariance matrix for each mass used in the fit at each point where the $\beta$ function was evaluated and proceeded as before.

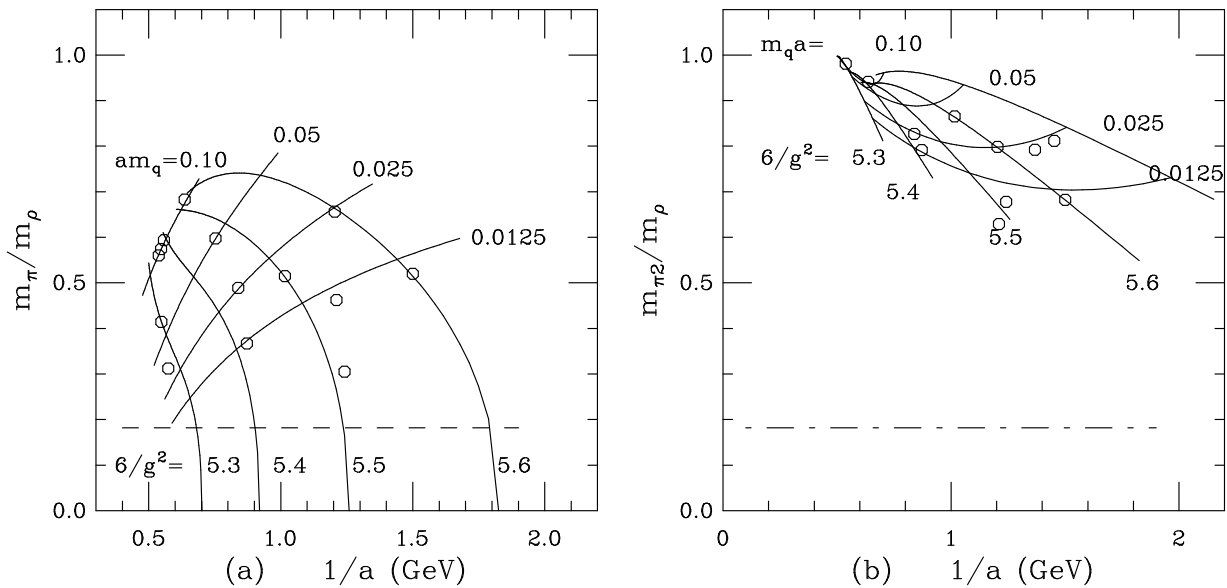

FIG. 4. Spectrum fit showing contours of constant bare coupling and quark mass. (a) Using the mass of the Goldstone pion of the exact $U(1) \times U(1)$ lattice chiral symmetry. (b) Using a non-Goldstone pion mass. The fitting functions are given in Table III. The dashed line is the physical value of $m_\pi/m_\rho$. The inverse function gives the renormalization group flow. The octagons indicate the points where zero temperature spectrum calculations were done. The solid lines are conours of constant bare parameters $6/g^2$ and $am_q$ with the values indicated in the graphs.

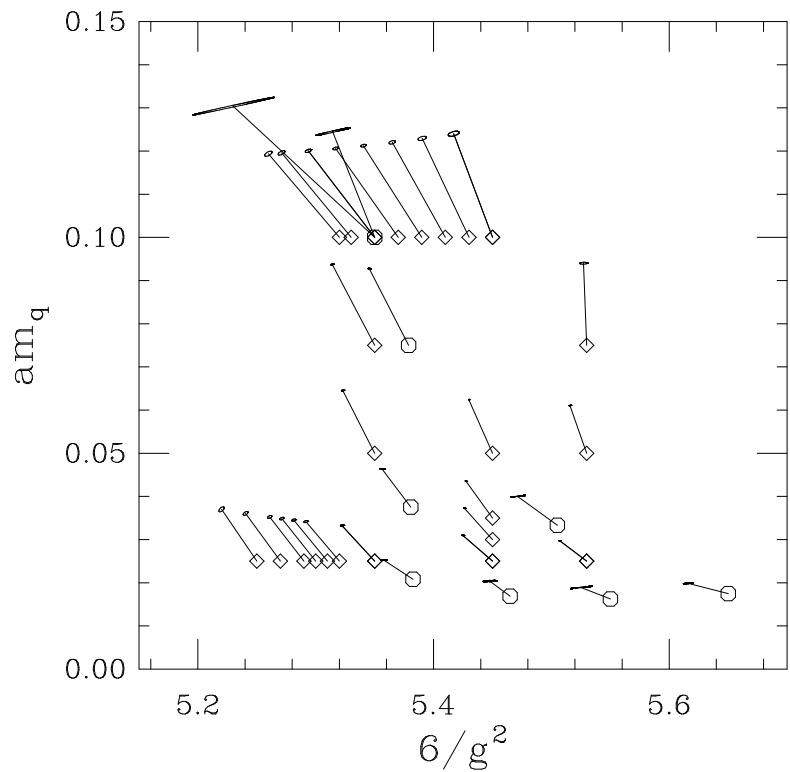

FIG. 5. Comparison of QCD nonperturbative $\beta$ function calculated directly from spectrum data and from the fit to the spectrum data. The octagons are the direct calculation, and the diamonds are from the fit. The locations of the octagons are averages of the points in parameter space used to calculate the $\beta$ function. The diamonds show where thermodynamic data were obtained for the equation of state.

## III. SIMULATIONS

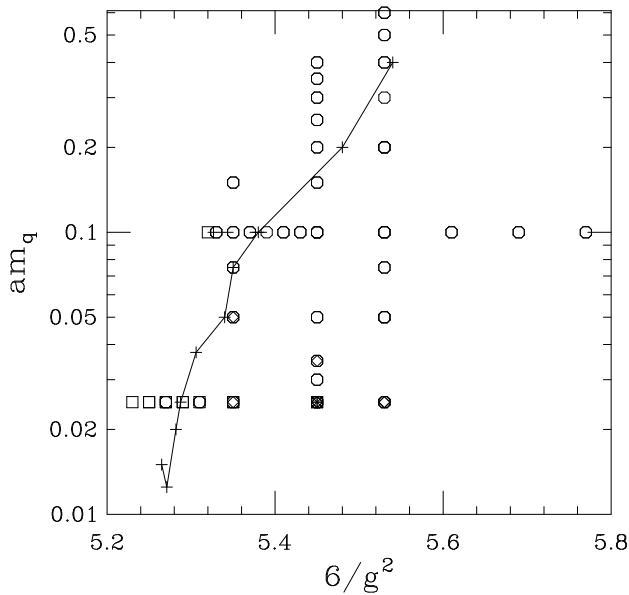

FIG. 6. The bare parameters of our runs. The plot symbols indicate the step size used, where the cross is $dt = 0.01$, the squares 0.02, the octagons 0.03, the diamonds 0.04 and the plusses 0.05. Some of the points have runs with several step sizes. The plusses connected by the solid line show estimates for the phase transition or crossover region for $N_t = 4$ [18,19].

We performed simulations with the parameter values displayed in Fig. 6. At each point we ran both hot ($N_t = 4$) and cold ($N_t = N_s$) lattices. For $6/g^2 \leq 5.45$ we used $N_s = 8$ or 12. For $5.45 < 6/g^2 \leq 5.69$ we used $N_s = 12$, and for $6/g^2 = 5.77$ we used $N_s = 16$. On the cold lattices we performed about 800 trajectories plus 100 warmup trajectories, on the hot runs about 1600 plus 100 warmup trajectories

As mentioned in Section 2, these results are subject to step size errors. We found that these are not the same for the hot and cold runs. In cold lattices the effect was much more pronounced. Thus it is not safe to subtract the results of cold and hot lattices without making sure that the step size errors are under control. This increases the workload considerably.

The system, with fixed step size $dt$ is self contained, that is it corresponds to some statistical system and the integrations along constant $am_q$ and $6/g^2$ should give consistent results also at non-zero $dt$, provided our integrations are numerically accurate. So, this self consistency, although not strictly speaking physical, can be used to test the accuracy of our integrations. Indeed at fixed step size, the integrations agreed within error bars.

In our case the proper handling of step size errors was especially important in the integration over $6/g^2$. This is because the difference of the hot and cold plaquettes was in many cases of the same order as the step size error. In Fig. 7a we show the plaquette as a function of step size squared, which is the leading error in the R-algorithm [7]. The difference in the step size errors between hot and cold runs also has implications for the operator measurements of pressure and energy density where a similar zero point subtraction is needed.

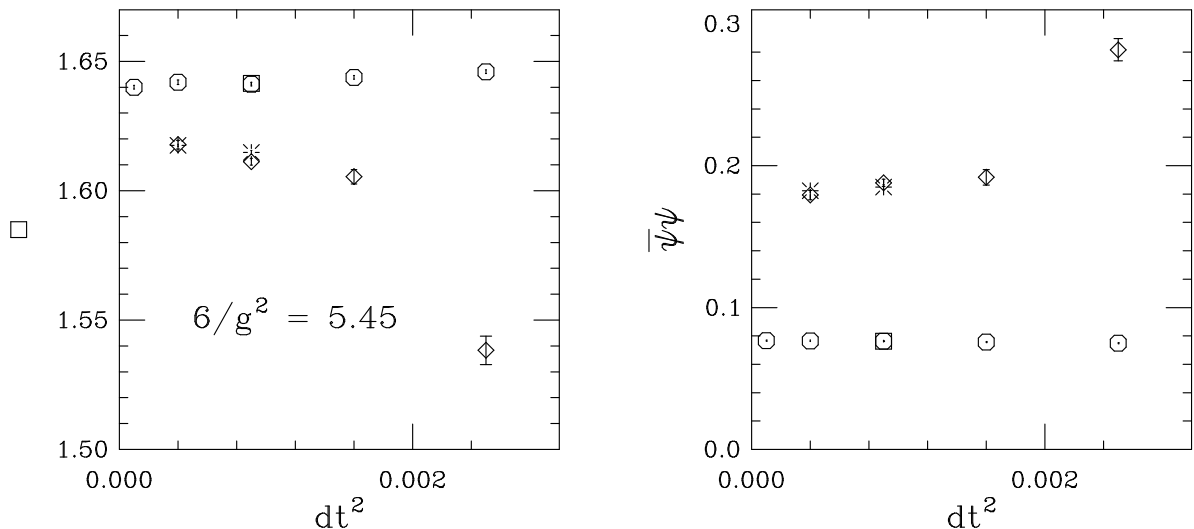

FIG. 7. (a) The variation of the plaquette as a function of step size squared at $6/g^2 = 5.45$ and $mq_a = 0.025$. The diamonds come from $8^4$ and the octagons from $8^3 \times 4$ lattices. The cold system has a much more pronounced effect. (b) The same for $\bar{\psi}\psi$. The cold system again has a larger effect, but it is much less significant for the calculation of the pressure, since the shift in $\bar{\psi}\psi$ due to the step size error is a smaller fraction of the difference between the hot and cold values. The squares and bursts are runs with $N_s = 12$.

For $\bar{\psi}\psi$ the step size error was not as large relative to the difference in values for the hot and cold condensate, as is seen in Fig. 7. Thus, the error in the measurement of the pressure from the $am_q$ integrations is smaller. However, for the smallest values, $6/g^2 = 5.35$ and $am_q = 0.025$, the extrapolation of $\bar{\psi}\psi$ to $dt = 0$ had to be taken into account: with step sizes 0.02, 0.03 and 0.04 the cold lattice $\bar{\psi}\psi$ was 0.3021(20), 0.3328(25) and 0.3867(33) respectively. Linear extrapolation in step size squared gave $\bar{\psi}\psi = 0.2731(28)$, significantly different from the smallest step size result. In contrast, the hot $\bar{\psi}\psi$ was almost the same at $dt = 0.02$ and 0.03. This was the worst case in the mass integrations.

For the $6/g^2$ integrations the step size errors are hardest to handle. The effects at $am_q = 0.1$ were within error bars. At $m_q a = 0.025$ there were two major effects. First, the cold plaquettes were smaller with larger step sizes, increasing the apparent difference between the hot and cold systems. This means, that at larger step sizes the pressure apparently becomes too large. Second, the position of the phase transition was shifted for the hot system towards smaller $6/g^2$ as step size is reduced. Therefore, the integrated pressure near the transition became smaller at larger step sizes, partially cancelling the first effect after integrating to large $6/g^2$. In all, the step size error caused the pressure to look steeper than it really is. In Fig. 8 we display the pressure for two different step sizes. To get the physical value we make a linear (in step size squared) extrapolation of the pressure to zero step size also shown in the figure. The extrapolation is in agreement with the values of pressure from the mass integrations. In our final results, when a point could be reached by two integration paths, we combined the results of the two paths with weights proportional to the inverse squares of their statistical errors. Roughly speaking, the effect of this on the pressure versus $6/g^2$ is to use the integrations over $am_q$, with smaller statistical errors, as "anchors" for the integrations over $6/g^2$.

Figures 9 and 10 collect our results for the pressure with $N_t = 4$. In Fig. 9 we have combined the results of integrating over $6/g^2$ and $am_q$.

Having the results as a function of $am_q$, we may wish to extrapolate to even smaller, physical quark masses. For small quark masses $am_q$ the $\bar{\psi}\psi$ should go as

$$\bar{\psi}\psi = h_0(6/g^2) + h_1(6/g^2) \cdot am_q + O(am_q)^2, T > 0$$

$$\bar{\psi}\psi = c_0(6/g^2) + c_1(6/g^2) \cdot am_q + O(am_q)^2, T = 0 \qquad (25)$$

where $h_i$ and $c_i$ are independent of mass, with $h_0 = 0$ for $T > T_C$. Using Eq. (12) one gets for the pressure

$$\begin{aligned}pa^4(am_q) &= (p_0 a^4) + \int_{m''}^{am_q} [(h_0 - c_0) + (h_1 - c_1)(am_q')] d(am_q') \\ &= (p_0 a^4) + (h_0 - c_0)(am_q - m'') + \frac{(h_1 - c_1)}{2}[(am_q)^2 - (m'')^2] + O(am_q^3), \qquad (26)\end{aligned}$$

where the lower limit $m''$ is small enough for Eq. (25) to be valid and $(p_0 a^4)$ is the value of the pressure at that point. Therefore, a simulation is needed only down to a mass value where the behavior of Eq. (25) is established.

However, if the transition is of second order, near the critical temperature the scaling of $\bar{\psi}\psi$ is governed by the critical exponent $\delta$ and these formulas must be altered. Using Eq. (26) to extrapolate the curves in Fig. 10 results in the bursts in Fig. 9.

It is interesting to notice that at high temperatures, the behavior of the pressure becomes ($h_0 \to 0$, $m'' \to 0$ )

$$pa^4(am_q) = p_0 a^4(m_q = 0) - \bar{\psi}\psi_{\text{cold}} m_q a^4 \qquad (27)$$

and its mass derivative is determined by the zero temperature $\bar{\psi}\psi$. In physical units

$$p(m_q) = p(0) - \langle\bar{\psi}\psi\rangle_{T=0} m_q \qquad (28)$$

Thus, even in very high energy scales, the zero temperature subtraction produces a nonzero derivative of the pressure with respect to quark mass at $am_q = 0$, in contrast to free quark behavior where this slope is zero.

The energy density from the interaction measure and $\beta$ function using Eqs. 13 and 14 is displayed in Figs. 11 and 12. The $\beta$ function was obtained from the fits to $m_\pi/m_\rho$ and $m_\rho$ in Table III.

The finite size error in the pressure comes from the assumption (Eq. 6) that the partition function scales with the volume. However, since most of the simulations were done in a region where the finite size effects in expectation values are small, we expect only small corrections. Even the integration through the crossover region is mostly at a large mass, where finite size effects are not so important. The correlation length is determined by the quark mass $am_q$ rather than by the lattice size $N_s$. The $6/g^2$ integration is over the crossover region at a small mass, where one might find an effect. But since the $6/g^2$ and $am_q$ integrations agree, the effect cannot be too large.

The largest systematic error comes from the fact that $N_t = 4$ lattices contain only a few of the Matsubara frequences in the partition function. One can estimate this by comparing the finite $N_t$ free field theory results to continuum results. The Boltzmann law for massless quarks in the continuum gives

$$\frac{\varepsilon}{T^4} = \frac{\pi^2}{30}(16 + 21) = 12.1725. \tag{29}$$

For free Kogut-Susskind quarks on a $12^3 \times 4$ lattice at $am_1 = 0.025$, the energy is

$$\frac{\varepsilon_{\text{gluon}}}{T^4}(N_t = 4, N_s = 40) = 1.46\frac{\pi^2}{30}16$$
$$\frac{\varepsilon_{\text{fermion}}}{T^4}(N_t = 4, N_s = 40) = 1.51\frac{\pi^2}{30}21 \tag{30}$$

giving $\varepsilon/T^4 = 18.1$. This is close to the value measured by us.

Fig. 4 shows how to translate our results to physical units (the coordinate axis can be changed to temperature by dividing by $N_t$). The lines of constant $am_q$ tend up as $6/g^2$ increases since the physical quark mass becomes larger as the lattice spacing $a$ is reduced at constant $am_q$. Fig. 4(a) shows that the most direct approach to the physical value of quark mass (or $m_\pi/m_\rho$), is to reduce $am_q$. However, Fig. 4(b) reminds us that at small $6/g^2$, even though the ratio of the Goldstone pion mass to the rho mass is near the physical value, the other pions are much heavier and the flavor symmetry is badly broken.

In Fig. 13 we show the pressure in physical units.

Finally, we look at the different contributions to the state variables from the fermionic and gluonic sectors. Since we have used the integration method for the pressure and did not directly measure the energy, we only look at the interaction measure. In Fig. 14 we show the contributions from both terms in Eq. 13 separately for the two lines of constant $m_q a$. Both pieces are normalized to zero at $T = 0$ by vacuum subtraction. As the transition region is traversed, the gauge piece shoots up while the fermion piece rises more slowly and more or less levels off. At high temperature the interaction measure is expected to go to zero. It is interesting to see how the two contributions approach this limit. In the high temperature phase, $\langle \bar{\psi}\psi \rangle$ approaches zero and the fermion contribution is given by the vacuum subtraction (a constant) times the quark mass component of the $\beta$ function which goes to zero in the continuum ($T \to \infty$) and chiral limits. Fig. 5 shows why this contribution is not falling off for $m_q a = 0.1$. The quark mass component of the $\beta$ function is actually increasing in this region of $6/g^2$ while for $m_q a = 0.025$ it is decreasing. The gauge contribution behaves in just the opposite way. The gauge component of the $\beta$ function goes to a constant while the average of the plaquette in the hot phase approaches the zero temperature value (see Fig. 1).

In this study we have developed and tested methods for determining the equation of state for high temperature QCD, and presented results with a large lattice spacing $a = 1/4T$. Simulations must be done at many values of $6/g^2$ and $am_q$, and extrapolations made to the physical values. Zero temperature simulations must be done to allow the divergent parts of the energy and pressure to be subtracted, to provide the $\beta$ function for computing the interaction measure, and to provide the mapping from the lattice variables to $6/g^2$ and $am_q$ to the physical variables $T$ and $m_\pi/m_\rho$. At the lattice spacing used here, lattice effects on the thermodynamics are large and flavor symmetry is strongly broken. We expect to pursue this project at smaller lattice spacing in the future.

## ACKNOWLEDGMENTS

These calculations were carried out on the Alpha cluster at the Pittsburgh Supercomputer Center (PSC) and the Paragons at the San Diego Supercomputer Center (SDSC) and Indiana University. We would like to thank Rob Pennington of PSC for his help with the Alpha cluster. Parts of this work were done at the Institute for Theoretical Physics in Santa Barbara. This research was supported by Department of Energy grants No. DE-FG02–85ER–40213 and DE-FG02–91ER–40661, and by the National Science Foundation under Grant No. PHY89-04035. LK wishes to thank the Antti Wihuri foundation for additional support.21

TABLE I. Spectrum runs used to construct the nonperturbative QCD $\beta$ function. We also tabulate the $\pi$ and $\rho$ masses, together with the values of our interpolating expression at these points, using the functions in Table III. Note that the points with $6/g^2 = 5.7$ were not used in the fitting.

| run | ref. | $am_q$ | $6/g^2$ | lattice size | $m_\pi$ | $m_\rho$ | $m_\pi^{fit}$ | $m_\rho^{fit}$ |
|---|---|---|---|---|---|---|---|---|
| 1 | [9] | 0.1 | 5.35 | $8^2 \times 24^2$ | 0.8019(1) | 1.432(5) | 0.805 | 1.438 |
| 2 | [10] | 0.1 | 5.375 | $8^3 \times 24$ | 0.8088(7) | 1.408(11) | 0.811 | 1.399 |
| 3 | [11] | 0.1 | 5.4 | $6^3 \times 24$ | 0.819(4) | 1.38(1) | 0.814 | 1.362 |
| 4 | [10] | 0.1 | 5.525 | $8^3 \times 24$ | 0.8262(8) | 1.210(16) | 0.817 | 1.211 |
| 5 | [10] | 0.05 | 5.32 | $8^3 \times 24$ | 0.5827(7) | 1.406(48) | 0.543 | 1.315 |
| 6 | [15] | 0.05 | 5.47 | $8^3 \times 24$ | 0.6112(5) | 1.023(2) | 0.611 | 1.024 |
| 7 | [10] | 0.025 | 5.2875 | $8^3 \times 24$ | 0.4184(9) | 1.342(46) | 0.375 | 1.274 |
| 8 | [12] | 0.025 | 5.445 | $16^3 \times 24$ | 0.4488(4) | 0.918(4) | 0.447 | 0.912 |
| 9 | [13] | 0.025 | 5.6 | $16^3 \times 32$ | 0.4197(7) | 0.6396(28) | 0.425 | 0.641 |
| 10 | [14] | 0.025 | 5.7 | $32^3 \times 32$ | 0.383(1) | 0.530(2) | 0.390 | 0.511 |
| 11 | [16] | 0.02 | 5.5 | $16^4$ | 0.3901(17) | 0.758(48) | 0.403 | 0.769 |
| 12 | [16] | 0.02 | 5.6 | $16^4$ | 0.3696(31) | 0.563(11) | 0.377 | 0.599 |
| 13 | [16] | 0.02 | 5.7 | $20^4$ | 0.3402(17) | 0.4916(30) | 0.338 | 0.464 |
| 14 | [12] | 0.0125 | 5.415 | $16^3 \times 24$ | 0.3239(5) | 0.883(6) | 0.330 | 0.891 |
| 15 | [16] | 0.01 | 5.5 | $16^4$ | 0.2942(44) | 0.636(37) | 0.293 | 0.692 |
| 16 | [13] | 0.01 | 5.6 | $16^3 \times 32$ | 0.2667(8) | 0.5133(22) | 0.263 | 0.512 |
| 17 | [16] | 0.01 | 5.7 | $20^3 \times 20^4$ | 0.2451(23) | 0.4184(70) | 0.219 | 0.368 |
| 18 | [17] | 0.004 | 5.48 | $16^3 \times 32$ | 0.189(1) | 0.62(4) | 0.197 | 0.686 |

TABLE II. Nonperturbative QCD $\beta$ function. The upper section is for the $\beta$ function calculated by the direct method. The first two entries are from new simulations and correspond to the VT $\rho$ and the PV $\rho$, respectively. In these new simulations the couplings were actually varied asymmetrically in space and time around the cited values. The other points are from the literature and correspond to the VT $\rho$. "runs" gives the runs from Table I used to construct the $\beta$ function. "corr" is the scaled correlation between the two components of the $\beta$ function. The lower section of the table contains the $\beta$ function obtained from the fits to the mass spectrum in Table III at a few selected points. Finally, we quote the perturbative value.

| $6/g^2$ | $am_q$ | runs | $\frac{\partial(6/g^2)}{\partial \ln(a)}$ | $\frac{\partial am_q}{\partial \ln(a)}$ | $\frac{\partial \ln(am_q)}{\partial \ln(a)} - 1$ | corr. |
|---|---|---|---|---|---|---|
| 5.35 | 0.1 | 1 | $-1.201(342)$ | 0.304(20) | 2.04(20) | $-0.99$ |
| 5.35 | 0.1 | 1 | $-0.379(128)$ | 0.253(8) | 1.53(8) | $-0.97$ |
| 5.379 | 0.075 | 1 2 5 6 | $-0.333(11)$ | 0.1774(9) | 1.37(1) | $-0.50$ |
| 5.381 | 0.0375 | 5 6 7 8 | $-0.243(15)$ | 0.0886(8) | 1.36(2) | $-0.77$ |
| 5.3825 | 0.0208 | 7 8 14 | $-0.251(23)$ | 0.0444(5) | 1.13(2) | $-0.84$ |
| 5.465 | 0.0169 | 8 11 14 15 | $-0.176(53)$ | 0.0351(15) | 1.08(9) | 0.40 |
| 5.505 | 0.0333 | 6 8 9 | $-0.338(58)$ | 0.067(2) | 1.01(6) | 0.98 |
| 5.55 | 0.01625 | 9 11 15 16 | $-0.249(88)$ | 0.027(3) | 0.64(18) | 0.98 |
| 5.65 | 0.0175 | 9 10 16 17 | $-0.344(35)$ | 0.024(2) | 0.37(11) | 0.84 |
| 5.35 | 0.1 | | $-0.565(25)$ | 0.200(4) | 1.00(4) | 0.52 |
| 5.35 | 0.025 | | $-0.278(14)$ | 0.082(2) | 2.28(8) | 0.09 |
| 5.55 | 0.025 | | $-0.226(3)$ | 0.0450(2) | 0.80(8) | 0.49 |
| 5.35 | 0.0 | | $-0.371(20)$ | - | - | - |
| 5.55 | 0.0 | | $-0.270(4)$ | - | - | - |
| $\to \infty$ | 0.0 | | $-0.734$ | - | - | - |

TABLE III. Fits to spectrum data used to construct the nonperturbative QCD $\beta$ function. (Not recommended for $6/g^2 > 5.6$)

$$m_\pi/m_\rho = \sqrt{am_q}(5.10 + 12.89(6/g^2 - 5.45) - 2.49(6/g^2 - 5.45)^2)$$
$$-am_q(15.05 + 34.38(6/g^2 - 5.45)) + 16.51(am_q)^{3/2}$$

$\chi^2$ with 8 degrees of freedom 23.4.

Confidence level 0.0029.

Runs from Table I used in fit $1, 2, 3, 4, 5, 6, 7, 8, 9, 11, 14, 15, 16, 18$

$$m_\rho = 0.72 - 2.25(6/g^2 - 5.45) + 1.75(6/g^2 - 5.45)^2 + 7.75am_q$$
$$+10.01am_q(6/g^2 - 5.45) - 20.08(am_q)^2$$

$\chi^2$ with 8 degrees of freedom 20.7.

Confidence level 0.008.

Runs from Table I used in fit $1, 2, 3, 4, 5, 6, 7, 8, 9, 11, 14, 15, 16, 18$

$$m_{\pi_2} = 0.49 - 2.20(6/g^2 - 5.45) + 3.00(6/g^2 - 5.45)^2 + 11.02am_q$$
$$+7.32am_q(6/g^2 - 5.45) - 36.61(am_q)^2$$

$\chi^2$ with 5 degrees of freedom 8.9.

Confidence level 0.11.

Runs from Table I used in fit $1, 4, 8, 9, 11, 12, 13, 14, 15, 16, 18$

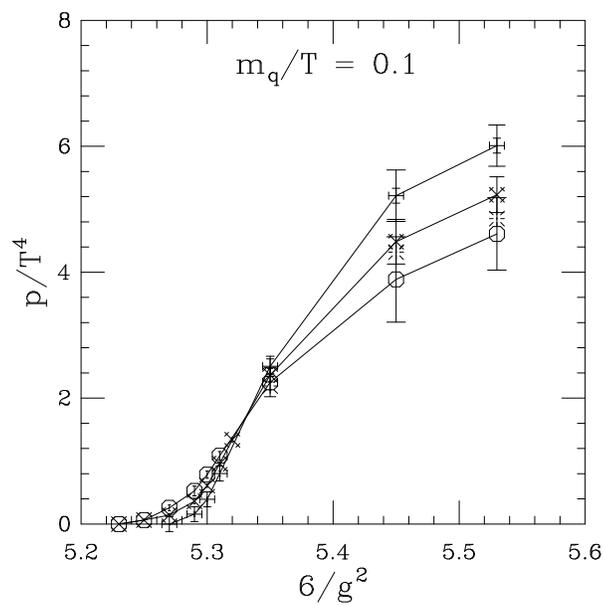

FIG. 8. The effect of the finite step size on the pressure. The fancy plusses are from step size 0.03, the fancy crosses from 0.02. The octagons give the result extrapolated to zero step size. The bursts show the mass integration results.

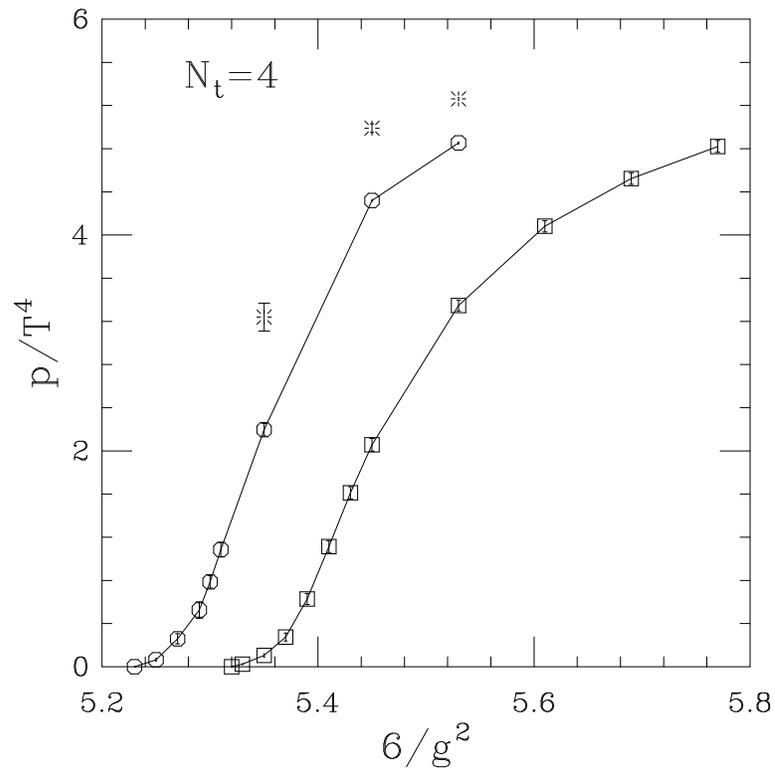

FIG. 9. The pressure in units of $T^4$ as a function of gauge coupling $6/g^2$ The bursts are extrapolations of the mass integrations to $am_q = 0$ (see Fig. 10 and Eq. 26).

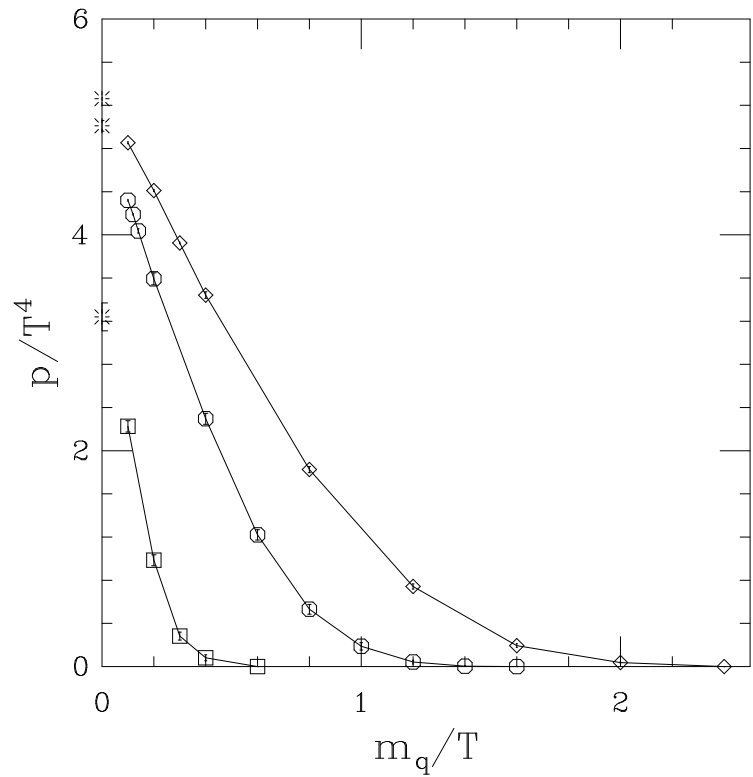

FIG. 10. The pressure in units of $T^4$ as a function of the quark mass $m_q/T$ from the integrations over mass. The diamonds show give $6/g^2 = 5.53$, the octagons $6/g^2 = 5.45$ and the squares the $6/g^2 = 5.35$ results. The bursts are extrapolations of the mass integrations to $am_q = 0$ (see Eq. 26).

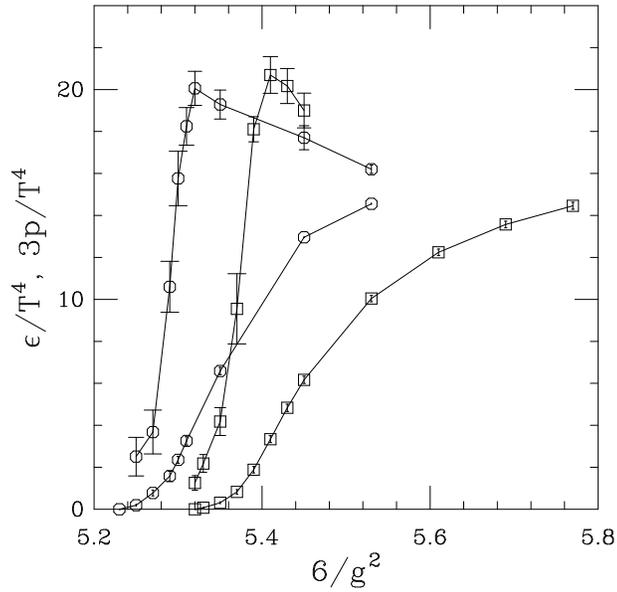

FIG. 11. The energy density in units of $T^4$ as a function of the gauge coupling. The lower curves are the pressure values displayed for comparison. The octagons are for $am_q = 0.025$, or $m_q = T/10$, and the squares are for $am_a = 0.1$, or $m_q = 0.4T$. The errors contain the uncertainty in the $\beta$ function. The $am_q = 0.1$ energy could not be computed at higher $6/g^2$ because of the lack of a reliable $\beta$ function in this region of $6/g^2$ and $am_q$.

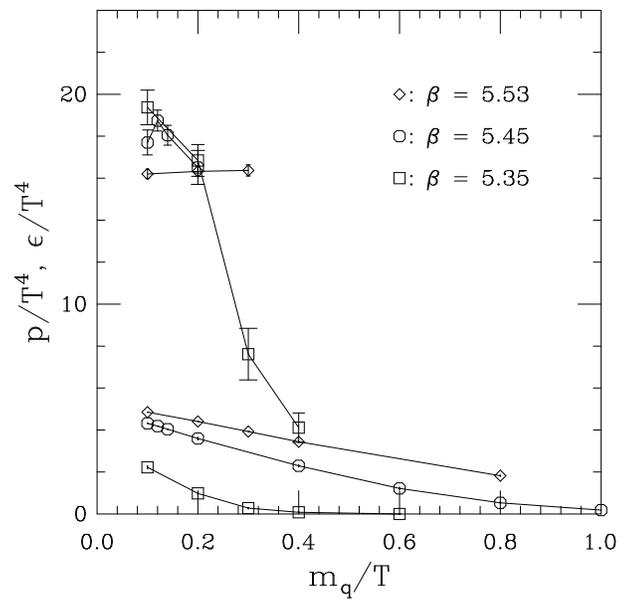

FIG. 12. The energy density in units of $T^4$ as a function of the quark mass $m_q/T$. Again, the pressure is also plotted.

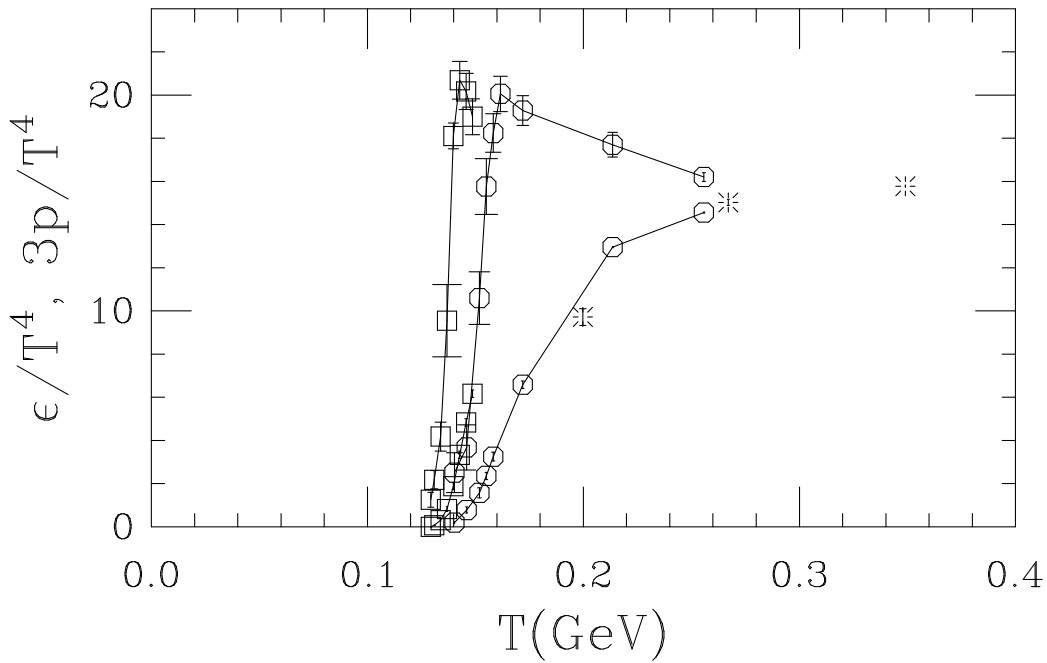

FIG. 13. The energy density in units of $T^4$ as a function of the temperature. The lower curves are three times the pressure displayed for comparison. The octagons are for $am_q = 0.025$, or $m_q = T/10$, and the squares are for $am_a = 0.1$, or $m_q = T/4$. The errors contain the uncertainty in the $\beta$ function. The $am_q = 0.1$ energy could not be computed at higher $6/g^2$ because of the lack of a reliable $\beta$ function in this region of $6/g^2$ and $am_q$. (The pressure curve for this case lies almost on top of the energy curve for $am_q = 0.025$ and is easily overlooked.) The bursts are extrapolations of the mass integrations to $m_q = 0$.

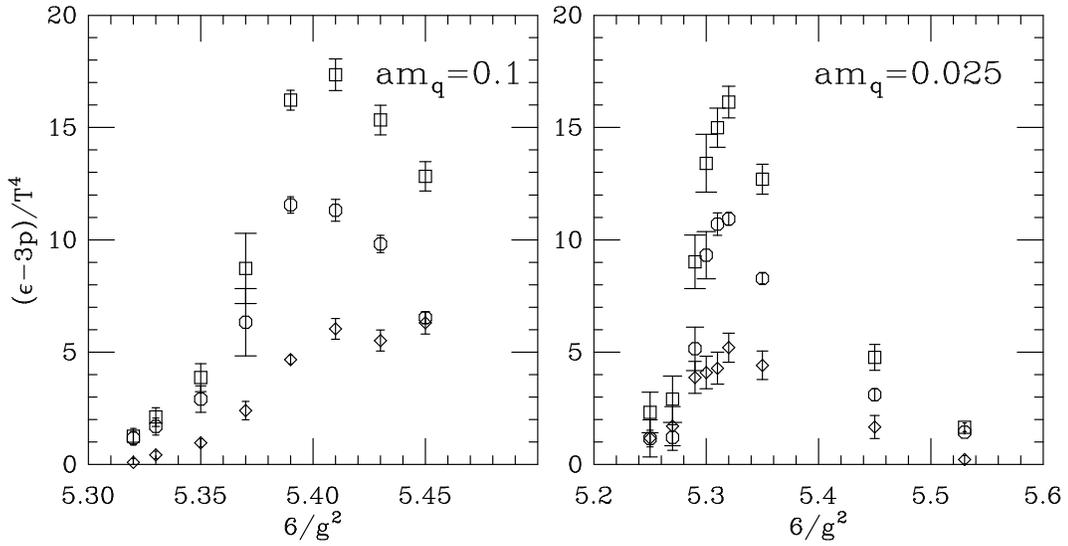

FIG. 14. Gauge and fermion contributions to the interaction measure for (a) $m_q a = 0.1$ and (b) $m_q a = 0.025$. The circles give the gauge part, and the diamonds give the fermion part. The squares are the total.